\documentclass[twocolumn,prl,aps,showpacs,superscriptaddress,amsmath,amssymb]{revtex4}

\usepackage{graphicx}
\usepackage{dcolumn}
\usepackage{bm}

\newcommand{\dbar}{{\bar{\partial}}}

\newcommand{\bfk}{{\mathbf{k}}}
\newcommand{\bfr}{{\mathbf{r}}}
\newcommand{\bfu}{{\mathbf{u}}}
\newcommand{\bfv}{{\mathbf{v}}}
\newcommand{\bfe}{{\mathbf{e}}}
\newcommand{\bfA}{{\mathbf{A}}}
\newcommand{\bfG}{{\mathbf{G}}}
\newcommand{\bfR}{{\mathbf{R}}}
\newcommand{\bfalpha}{{\boldsymbol{\alpha}}}

\newcommand{\ee}{{\mathrm{e}}}
\newcommand{\dd}{{\mathrm{d}}}
\newcommand{\tr}{{\mathrm{tr}}}
\newcommand{\Tr}{{\mathrm{Tr}}}

\newcommand{\WS}{{C}}
\newcommand{\BZ}{{\Omega}}
\newcommand{\BZC}{{\Omega_\mathbb{C}}}

\begin{document}

\preprint{APS/123-QED}

\title{Exponential localization of Wannier functions in insulators}%

\author{Christian Brouder}
\affiliation{%
Institut de Min\'eralogie et de Physique des Milieux Condens\'es,
CNRS UMR 7590, Universit\'es Paris 6 et 7, IPGP, 140 rue de Lourmel,
75015 Paris, France.
}%
\author{Gianluca Panati}
\affiliation{%
Zentrum Mathematik and Physik Department,
Technische Universit{\"a}t M{\"u}nchen, 80290 M{\"u}nchen, Germany.
}%
\author{Matteo Calandra}
\affiliation{%
Institut de Min\'eralogie et de Physique des Milieux Condens\'es,
CNRS UMR 7590, Universit\'es Paris 6 et 7, IPGP, 140 rue de Lourmel,
75015 Paris, France.
}%
\author{Christophe Mourougane}
\affiliation{%
Institut de Math{\'e}matiques de Jussieu,
175 rue du Chevaleret, 75013 Paris, France.
}%
\author{Nicola Marzari}
\affiliation{%
Department of Materials Science and Engineering,
Massachusetts Institute of Technology,
77 Massachusetts Avenue, Cambridge MA 02139-4307, USA.
}%

\date{\today}

\begin{abstract}
The exponential localization of Wannier functions in two or three dimensions
is proven for all insulators that display time-reversal symmetry,
settling a long-standing conjecture.  Our proof relies on the equivalence 
between the existence of analytic quasi-Bloch functions and
the nullity of the Chern numbers (or of the Hall current) for the system 
under consideration. 
The same equivalence implies that Chern insulators cannot display 
exponentially localized 
Wannier functions.  An explicit condition for the reality of the 
Wannier functions is identified.
\end{abstract}

\pacs{71.23.An, 03.65.Vf}
\maketitle


Wannier functions play a fundamental role in the 
description of the electronic properties of solids \cite{Kohn64}.
They allow for an intuitive interpretation of the
bonding properties of solids \cite{Marzari97}, they are at the center
of the modern theory of polarization \cite{Vanderbilt93}, and they form
a very efficient basis for order-$N$ calculations or
the construction of model Hamiltonians \cite{Marzari05}.

For one-dimensional systems, Kohn \cite{Kohn59} proved that
Wannier functions are exponentially localized.
This property has many desirable consequences, 
such as the existence of moments $\langle r^n\rangle$
for all $n$, the exponential
convergence of numerical calculations \cite{Stengel},
and the possibility to use Wannier functions for the
description of surfaces \cite{RehrKohn}.
In two and three dimensions, the existence of
exponentially localized Wannier functions is one
of the few unsolved problems of one-particle
condensed-matter physics \cite{Nenciu83}.
In the absence of a proof for exponential decay,
this property is commonly used as working
hypothesis \cite{Prodan}, checked with numerical calculations
\cite{Schnell}, or simply taken for granted \cite{KimMauri,Nunes}.

In this paper, we demonstrate that exponentially localized Wannier 
functions exist for insulators in two and three dimensions, and show how 
localization is related to the Berry connection for the set of bands 
under consideration \cite{Marzari97} and to the corresponding Chern number(s).
Chern numbers have been playing a rapidly increasing role in 
solid-state physics, from metal-insulator transitions \cite{ShengWeng}, to
quantized transport \cite{WalkerWilkinson},
transition metal nanomagnets \cite{Canali},
quantum Hall effect \cite{Hatsugai}, and its spin analogue
\cite{ShengHaldane}. Our central result is that
Wannier functions with exponential decay can
be constructed if and only if all the Chern numbers are zero.
This implies that if the system is symmetric for time-reversal,
then the Wannier functions are exponentially localized. 

We consider a crystal with a Bravais lattice $\Gamma$
and a unit cell $\WS$. The reciprocal lattice is
denoted by $\Gamma^*$ and the reciprocal unit cell
by $\BZ$. A function $f(\bfr)$ is 
called \emph{periodic} if $f(\bfr+\bfR)=f(\bfr)$ for any 
vector $\bfR$ of $\Gamma$, while a function $f(\bfk)$ is 
called \emph{periodic in the reciprocal space} if $f(\bfk+\bfG)=f(\bfk)$ 
for any vector $\bfG$ of $\Gamma^*$.
The dynamics of the electrons in the crystal is described by
a Hamiltonian $H = -\Delta + V(\bfr)$ (in Ry),
where the potential $V$ is real and periodic.
According to Bloch's theorem, the eigenfunctions 
of $H$ can be written
as $\psi_{n\bfk}(\bfr)=\ee^{i\bfk\cdot\bfr}u_{n\bfk}(\bfr)$,
where the Bloch states $u_{n\bfk}(\bfr)$ are periodic eigenstates of the 
Hamiltonian $H(\bfk)=(-i\nabla+\bfk)^2 +V(\bfr)$ with
energy $\epsilon_n(\bfk)$ and satisfy the boundary conditions
\begin{eqnarray}
u_{n\bfk+\bfG}(\bfr) &=& \ee^{-i\bfG\cdot\bfr}u_{n\bfk}(\bfr).
\label{condlim}
\end{eqnarray}

Wannier functions are defined as
\begin{eqnarray}
w_{n}(\bfr-\bfR) &=& \frac{1}{|\BZ|}
\int_{\BZ} \dd\bfk \ee^{i\bfk\cdot(\bfr-\bfR)} u_{n\bfk}(\bfr),
\label{defWannier}
\end{eqnarray}
where $|\BZ|$ is the volume of $\BZ$. 
The localization properties of the Wannier functions are related
to the regularity of $u_{n\bfk}$ as a function of $\bfk$. 
In a nutshell, the more regular the states, the more localized
the Wannier functions \cite{Kohn59,Cloizeaux1,Strinati78}. 
Exponential decay is obtained if and
only if the functions are analytic 
\cite{Cloizeaux1,Katznelson}.

In the simplest procedure,
the energies $\epsilon_n(\bfk)$
and the Bloch functions $u_{n\bfk}$ are determined
as a function of $\bfk$
by ordering the eigenvalues by increasing energies
$\epsilon_1(\bfk) \le \epsilon_2(\bfk) \le \dots$. 
Although this procedure is standard in band-structure calculations,
it gives Bloch states $u_{n\bfk}$ that are not regular
in $\bfk$ since the phase of $u_{n\bfk}$ is random. Moreover,
at a crossing point the energy $\epsilon_n(\bfk)$ can have a kink. 
In one dimension, Kohn \cite{Kohn59} showed that it is possible
to define Bloch states that are analytic functions of $\bfk$.
In two and three dimensions this is generally not possible \cite{Cloizeaux1}
because band crossings lead to Wannier functions that 
can decay as $1/R^4$ \cite{Bross}.

\begin{figure}
\includegraphics[width=8.0cm]{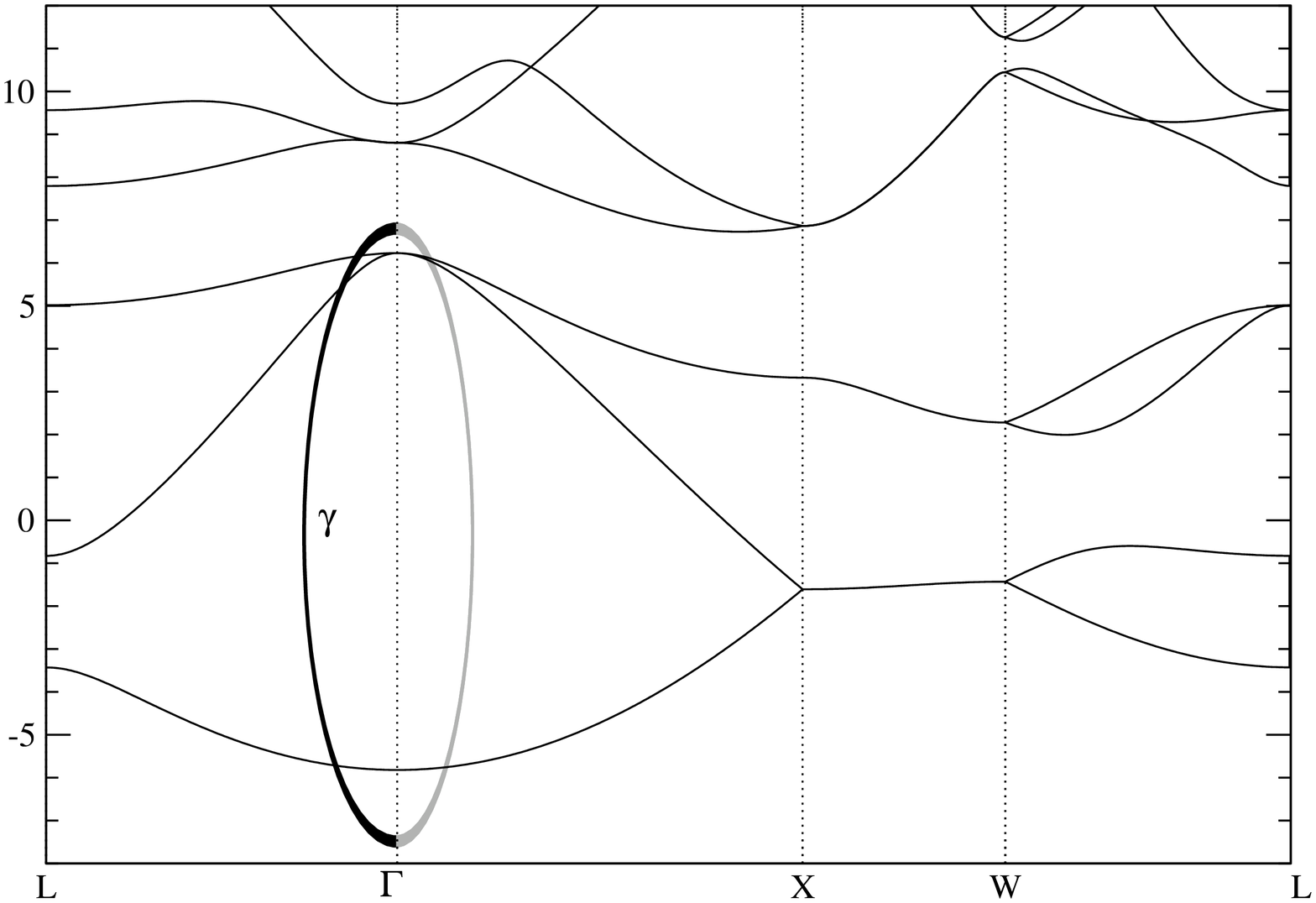}
\caption{Band structure of silicon. The contour $\gamma$
  encloses the energies of the composite band. \label{fig:Si}}
\end{figure}

Blount \cite{Blount} noticed that the decay properties
of Wannier functions can be improved by considering
a set of eigenstates, called  a \emph{composite band} \cite{Kohn73},
which are seaparated by a gap from all others.
More precisely, if $\epsilon_1(\bfk),\dots,\epsilon_M(\bfk)$
are the eigenstates of the composite band, we assume that
there is an $a>0$ such that $|\epsilon_n(\bfk)-\epsilon_m(\bfk)|>a$
for any $\bfk$ in $\BZ$, $1\le n\le M$ and $m >  M$.
In insulators it is always possible to define a composite band
since gaps exist between valence and conduction bands and
between valence bands and core states. This is not necessarily
the case in metals.  An example of a composite band 
is given in Fig.\ref{fig:Si} for silicon. 
Exponentially localized Wannier functions are obtained if
we can find
$M$ analytic functions $v_{n\bfk}$ (called \emph{quasi-Bloch}
states) that span the same vector space as the 
$M$ Bloch states of the composite band.
The problem of the existence of analytic quasi-Bloch states
was solved for a single isolated band 
(i.e. $M=1$) in 3D by des Cloizeaux \cite{Cloizeaux1}
and Nenciu \cite{Nenciu83}.
In this work we determine when analytic quasi-Bloch functions exist
for any composite band in two and three dimensions.

The quasi-Bloch functions $v_{n\bfk}$ can be expressed
as $v_{n\bfk}=\sum_m u_{m\bfk}(\bfr) U_{mn}(\bfk)$, where
$U(\bfk)$ is a unitary matrix, defining the 
Wannier functions
\begin{eqnarray}
w_{n}(\bfr) &=& \frac{1}{|\BZ|}
\int_{\BZ} \dd\bfk \ee^{i\bfk\cdot\bfr} \sum_m 
u_{m\bfk}(\bfr) U_{mn}(\bfk).
\label{defGWannier}
\end{eqnarray}
The improved localization of these Wannier functions was
observed on many systems \cite{Teichler,Marzari97}.

To study the analytic properties of the Bloch functions,
we consider a complex vector ${\bf k}={\bf k'}+i{\bf k''}$.
Then, $H(\bfk)$ is not Hermitian
but there still are eigenvalues $\epsilon_n(\bfk)$ and eigenstates
$u_{n\bfk}$ such that
$H(\bfk)u_{n\bfk}=\epsilon_n(\bfk) u_{n\bfk}$.
The branch points of $\epsilon_n(\bfk)$
determine the points of non-analyticity of the function
$u_{n\bfk}$ \cite{Blount,Kohn59}. 
Even very simple crystals, such as silicon,
exhibit such branch points \cite{ChangSi}.

If the potential $V(\bfr)$ is square integrable,
it can be shown that
$H(\bfk)$ is analytic \cite{RSIV}. 
The reality of $V(\bfr)$ implies that $H^\dagger(\bfk)=H(\bfk^*)$ and 
$H^*(\bfk)=H(-\bfk^*)$. 
From a practical point of view, the condition of
square integrability of $V(\bfr)$ encompasses potentials with
Coulomb singularities and the potentials
used in LDA and GGA calculations. 
The calculation of band structures in the complex
plane is available in some standard band-structure packages 
\cite{Smogunov}.

Several authors \cite{Cloizeaux1,Nenciu83,NenciuRMP} noted
that the obstacles to
the existence of quasi-Bloch states are topological.
To illustrate this, we introduce a
simple mathematical concept, that of a \emph{fibre}.
It is quite common in physics to consider a space
(let us call it the \emph{base}) and to associate
a vector space (called the \emph{fibre}) to each
point of the base. For example, in magnetostatics,
the base $B$ is the space $\mathbb{R}^3$ and to each point
$\bfr$ of the base we associate a three-dimensional
vector space (the fibre $F(\bfr)=\mathbb{R}^3$).
The vector potential $\bfA(\bfr)$ is a vector of the
space $F(\bfr)$. In the Born-Oppenheimer
approximation, the base $B$ is the set of possible positions
of the nuclei $\bfR_1,\dots,\bfR_N$ and the fiber 
$F(\bfR_1,\dots,\bfR_N)$
corresponding to a nuclear configuration is the vector
space generated by the solutions of the Schr\"odinger
equation with clamped nuclei. 
For a composite band, the base is the Brillouin zone 
$\BZ$ and the fibre $F(\bfk)$ is the $M$-dimensional vector space 
generated by the Bloch states $u_{n\bfk}$.
The quasi-Bloch states $v_{n\bfk}$ can now be defined precisely 
as a basis of $F(\bfk)$ such that each
$v_{n\bfk}$ is analytic in $\bfk$, periodic in the reciprocal
space and satisfies the boundary conditions (\ref{condlim}).

To show that topology might forbid the existence of
quasi-Bloch states,
we consider the simple example where the base $B$ is a circle
represented by the angular variable $x\in [0,2\pi]$ 
and the fibre is $F(x)=\mathbb{R}$. 
\begin{figure}
\includegraphics[width=8.0cm]{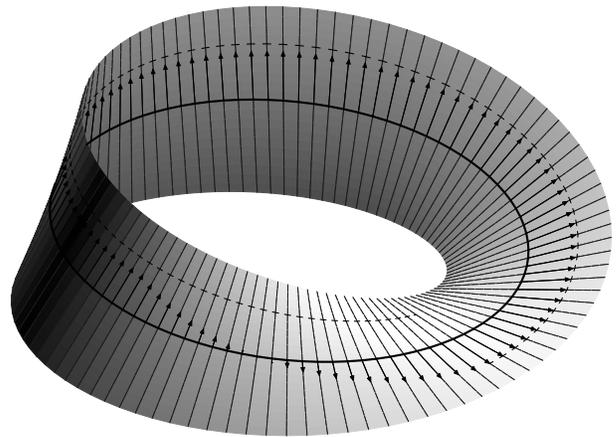}
\caption{M{\"o}bius strip. The solid circle is the base $B$
  and the thin straight lines are the fibres $F(x)$. 
  The dashed line represents the constant
  but not antiperiodic basis $\bfe$.
  The arrows represent the continuous and antiperiodic basis
  vectors $\bfv(x)$. It is clear that $\bfv(x)$ has to be zero
  for some $x$. \label{fig:Moebius}}
\end{figure}
We denote by $\bfv(x)$ a basis of $F(x)$.
Topology intervenes when we determine how the 
fiber at $x=0$ is related to the fiber at $x=2\pi$:
we can use periodic boundary conditions 
$\bfv(2\pi)=\bfv(0)$ (cylinder) 
or antiperiodic boundary conditions 
$\bfv(2\pi)=-\bfv(0)$ (M{\"o}bius strip,
illustrated in Fig. 2).  
In the case of a cylinder, we can choose a basis
$\bfv(x)={\bf e}$ (a non-zero vector independent of $x$).
This basis is obviously regular and periodic in $x$.
In the case of a M{\"o}bius strip, let us assume
that a regular and antiperiodic basis ${\bf v}(x)$ of $F(x)$ 
exists. Let $y(x)$ be the component
of ${\bf v}(x)$ with respect to the periodic basis $\bfe$,
namely ${\bf v}(x)=y(x)\bfe$. 
Clearly $y(x)$ has to be regular and antiperiodic
($y(0)=-y(2\pi)$).
By the intermediate value theorem
there needs to be a point $x_0$ between $0$ and $2\pi$
such that $y(x_0)=0$, as shown in 
Fig.~\ref{fig:Moebius}. Thus at $x_0$, ${\bf v}(x_0)={\bf 0}$.
But the null vector cannot be a basis for $F(x_0)$. Therefore, 
the topology of the M\"obius strip implies that
no regular basis of $F(x)$ can exist.

The topological obstruction to the existence of
quasi-Bloch states in dimensions one, two and three
has been recently studied \cite{Panati}. It was
discovered that quasi-Bloch states exist if and only
if all the Chern numbers of the system are zero. Such result
is of central importance to our paper, since it shifts the
focus of our analysis on the determination of the nullity of
Chern numbers in dimensions two and three.

For a given ${\bf k}$,
$P(\bfk)=\sum_{n=1}^{M} |u_{n\bfk}\rangle\langle u_{n\bfk^*}|$
defines a projector \cite{Blount}.
The Riesz formula for the projector is
\begin{eqnarray*}
P(\bfk) &=& \frac{1}{2\pi i} \int_\gamma \frac{\dd z}{z-H(\bfk)}.
\end{eqnarray*}
where $\gamma$ is a contour enclosing all the eigenvalues
$\epsilon_{n{\bfk}}$ of the composite band (see fig.\ref{fig:Si}).
$P(\bfk)$ is analytic in $\bfk$ on a strip 
$\BZC=\{\bfk=\bfk'+i\bfk'', \bfk' \in \BZ, |{k_j''}|<A,
j=1,2,3\}$ with $A>0$ even 
if the states $u_{n\bfk}$ are not analytic in ${\bfk}$
\cite{Cloizeaux1}.
The value of $A$ is related to the band gap \cite{Goedecker99}.
An example of contour $\gamma$ for Si is given in Fig. \ref{fig:Si}.

To calculate the Chern numbers, we introduce
the Berry connection corresponding to the basis 
$u_{n\bfk}$ of $F(\bfk)$, defined in dimension $d$
as the $d$-dimensional vector
of matrix functions $\bfA(\bfk)=\big(A^1(\bfk),\dots,A^d(\bfk)\big)$ 
with matrix elements
\begin{eqnarray*}
\bfA_{mn}(\bfk) &=& \int_C 
   u_{m\bfk}^*(\bfr) \nabla_\bfk u_{n\bfk}(\bfr)\dd\bfr.
\end{eqnarray*}
The trace of the curvature of this connection is \cite{Marzari97,Panati}
\begin{eqnarray}
B^{ij}(\bfk) &=& 
\tr\Big(\frac{\partial A^j}{\partial k_i}-
   \frac{\partial A^i}{\partial k_j}+
  {[}A^i,A^j{]}\Big)
\label{courbure}
\\&=&
  \Tr \Big( P(\bfk)
   \Big{[} \frac{\partial P(\bfk)}{\partial k_i},
   \frac{\partial P(\bfk)}{\partial k_j} \Big{]}\Big),
\nonumber
\end{eqnarray}
where $\tr$ is the matrix trace and $\Tr$ the operator trace.

In two dimensions, this curvature leads to the 
unique Chern number
\begin{eqnarray}
C_1 &=& \frac{i}{2\pi} \int_{\BZ} B^{12}(\bfk)
  \dd k_1 \dd k_2.
\label{Chern2D}
\end{eqnarray}
In three dimensions, we have three Chern-like numbers
\begin{eqnarray}
C_\ell &=& \frac{i}{2\pi} \sum_{i<j}\int_{\Theta_\ell} B^{ij}(\bfk)
  \dd k_i \wedge \dd k_j,
\label{Chern3D}
\end{eqnarray}
where $\ell=1,2,3$. The domain of integration $\Theta_\ell$
is a torus defined as the set of points $(k_1,k_2,k_3)$ of $\BZ$ 
such that $k_\ell=0$.
The reality of $V(\bfr)$ implies the time-reversal symmetry
$P(\bfk) = \big(P(-\bfk)\big)^*$. Therefore,
$B^{ij}(-\bfk)=-B^{ij}(\bfk)$, the Chern numbers are zero
and quasi-Bloch functions exist.

With this result in hand, we can repeat
the reasoning of des Cloizeaux (section III.B of 
Ref.~\onlinecite{Cloizeaux1}),
to show that
$\lim_{|R|\rightarrow \infty} \ee^{b|R|}w_n(\bfr-\bfR) = 0$
for any $b<A$.

Our proof was given for the case of ``spinless" electrons.
To take spin, spin-orbit and all relativistic corrections
into account, we consider a crystal described by
the Dirac Hamiltonian \cite{Strange}
$H(\bfk)=-ic\bfalpha\cdot(\nabla+i\bfk)+\beta c^2/2 + V(\bfr)$.
The Bloch functions are Dirac spinors $u^\alpha_{n\bfk}(\bfr)$
and the corresponding Berry connection is
\begin{eqnarray*}
A^{\alpha\beta}_{mn}(\bfk) &=&
\int \dd \bfr 
\big(u^\alpha_{m\bfk}(\bfr)\big)^* \,\nabla_\bfk u^{\beta}_{n\bfk}(\bfr).
\end{eqnarray*}
The curvature is defined by equation (\ref{courbure}),
where the trace is over the $n$ and $\alpha$ indices,
and the Chern numbers by eqs.(\ref{Chern2D})
and (\ref{Chern3D}).
Again, quasi-Bloch functions
exist for the Dirac Hamiltonian if and only if
all Chern numbers are zero \cite{Panati}
(in particular, if the potential $V$
is square integrable and time-reversal symmetric).
In that case, the
relativistic Wannier functions are exponentially localized.

Given the existence of
exponentially decaying Wannier functions $v_{n\bfk}$,
we need an algorithm to determine 
the unitary matrix $U(\bfk)$ such that
$v_{n\bfk}(\bfr)=\sum_{m} U_{mn}(\bfk)u_{m\bfk}(\bfr)$.
We sketch now a possible approach.
Being analytic, the quasi-Bloch functions satisfy the
Cauchy-Riemann equation
$\dbar v_{n\bfk}=0$, where $\dbar=(\dbar_x,\dbar_y,\dbar_z)$
and, if $\bfk=\bfk'+i\bfk''$,
$\dbar_j=(1/2)(\partial_{k'_j}+i\partial_{k''_j})$.
If we denote by 
$\bfu(\bfk)$ and $\bfv(\bfk)$ the vectors with components $u_{n\bfk}$
and $v_{n\bfk}$, the Cauchy-Riemann equation gives us
$\dbar\bfv=(\dbar \bfu) U + \bfu (\dbar U)=0$.
Multiplying by the conjugate of $\bfu$, we find
$\dbar U= X U$, where
$X_{mn}(\bfk) = -\langle u^*_{m-\bfk}|\dbar u_{n\bfk}\rangle$.
Note that, on regions where $u_{n\bfk}$ is analytic,
$X_{mn}(\bfk)=0$.
The equation $\dbar U=XU$ should be solved in the 
space of matrices $U(\bfk)$ periodic in the reciprocal space
such that $U(\bfk)^\dagger=U(\bfk^*)^{-1}$.
This condition ensures the orthogonality
of the quasi-Bloch functions on the real axis.
Explicit integral expressions are available to solve 
the Cauchy-Riemann equation $\dbar f =g$ \cite{Krantz1}.
They turn $\dbar U=XU$ into an integral
equation that can be solved numerically.

Note that, if $U$ is a particular solution of $\dbar U=XU$,
the general solution can be written
$U A$, where $A$ is any analytic matrix periodic in the
reciprocal space and satisfying $A(\bfk)^\dagger=A(\bfk^*)^{-1}$.
For instance, $A(\bfk)$ can be any matrix of the form
$A(\bfk)=\exp\big(\sum_\bfR a_\bfR \ee^{i\bfk\cdot\bfR}\big)$,
where the sum is over a finite number of sites of 
$\Gamma$ and the matrices $a_\bfR$ satisfy
 $a_\bfR^\dagger=-a_{-\bfR}$.
This very large set of solutions corresponds to
the fact that exponential decay is a long-range
property: Any finite linear combination
of exponentially decaying Wannier functions centered
on various sites is still 
exponentially decaying. Therefore, 
it is still necessary to optimize localization 
around the centers of the Wannier functions
\cite{Marzari97}
by properly choosing $A(\bfk)$. Such an approach
provides Wannier functions that are localized
on the short and long range.

It is convenient to determine under which
condition the Wannier functions are real.
In eq. (\ref{defGWannier}), we take the complex
conjugate and change the variables $\bfk\rightarrow -\bfk$
\begin{eqnarray*}
w_{n}^*(\bfr) &=& \frac{1}{|\BZ|} \sum_{m}
\int_{\BZ} \dd\bfk \ee^{i\bfk\cdot\bfr}
  u_{m-\bfk}^*(\bfr)(U_{mn}(-\bfk))^*.
\end{eqnarray*}
The reality of $V(\bfr)$ implies that
we can choose
$u_{m-\bfk}^*(\bfr)=u_{m\bfk}(\bfr)$.
Thus, the Wannier functions are real if
$(U_{mn}(-\bfk))^*=U_{mn}(\bfk)$ for real
$\bfk$. This property is satisfied if
$U$ satisfies
$U(-\bfk^*)=U(\bfk)^*$ for complex $\bfk$.

Wannier functions are often obtained by minimizing a functional
$\Omega(U)$ \cite{Marzari97}.
The time-reversal symmetry $I$ acts as
$(IU)_{mn}(\bfk)=(U_{mn}(-\bfk))^*$
for real $\bfk$.
If the functional $\Omega$ satisfies the symmetry
$\Omega(IU)=\Omega(U)$ and if 
$\Omega$ has a unique minimum (up to a possible 
overall phase),
then $U=IU$ and the Wannier functions are real.
The spread functional
$\Omega$ defined by  Marzari and Vanderbilt \cite{Marzari97}
satisfies the symmetry $\Omega(IU)=\Omega(U)$.
Thus, when the spread functional has a unique minimum $U$,
the corresponding Wannier functions are real, proving for this
case the conjecture of Ref.~\onlinecite{Marzari97}.

In this paper, we assumed that the system is
time-reversal symmetric. If this is not the case
(e.g. for the Haldane Hamiltonian \cite{Haldane}), the present
approach implies that exponentially localized
Wannier functions exist in regions of the parameter space
where the Chern numbers are zero or, equivalently, when
the Hall current is zero. Thus, the vanishing of the
Hall current is a measure of the exponential localization
of the Wannier functions. This confirms rigorously 
Thouless' observation \cite{Thouless}.
As a corollary, we deduce 
that no exponentially localized Wannier functions exist for
Chern insulators (i.e. insulators with non-zero Chern numbers
\cite{Ohgushi}) and that time-reversal symmetric systems
cannot be Chern insulators (as noticed by Haldane
\cite{Haldane} for the case $M=1$).

In conclusion, in this work we demonstrated that Wannier functions
are exponentially localized for insulators that satisfy time-reversal symmetry,
and we showed that the vanishing of the Chern numbers is equivalent
to the exponential localization of the Wannier functions.
As a corollary, Wannier functions in Chern insulators are not
exponentially localized. Moreover, we presented a simple criterion to determine
when Wannier functions can be chosen as real.

Electron localization is the key of several physical concepts
such as electric polarization \cite{RestaRMP}, piezoelectricity, orbital
magnetization \cite{Thonhauser} and the nature of the
insulating state \cite{Kohn64,Prodan}.  Our condition for
the occurrence of localization (the vanishing of Chern numbers)
is consequently a fundamental result for all these
subjects.

We thank K. K. Uhlenbeck, M. Putinar, R. Zentner,
D. S. Freed, F. Mauri and D. Ceresoli for useful comments,
and IDRIS (project 061202) for computer time.


\begin{thebibliography}{36}
\expandafter\ifx\csname natexlab\endcsname\relax\def\natexlab#1{#1}\fi
\expandafter\ifx\csname bibnamefont\endcsname\relax
  \def\bibnamefont#1{#1}\fi
\expandafter\ifx\csname bibfnamefont\endcsname\relax
  \def\bibfnamefont#1{#1}\fi
\expandafter\ifx\csname citenamefont\endcsname\relax
  \def\citenamefont#1{#1}\fi
\expandafter\ifx\csname url\endcsname\relax
  \def\url#1{\texttt{#1}}\fi
\expandafter\ifx\csname urlprefix\endcsname\relax\def\urlprefix{URL }\fi
\providecommand{\bibinfo}[2]{#2}
\providecommand{\eprint}[2][]{\url{#2}}

\bibitem[{\citenamefont{Kohn}(1964)}]{Kohn64}
\bibinfo{author}{\bibfnamefont{W.}~\bibnamefont{Kohn}}, \bibinfo{journal}{Phys.
  Rev.} \textbf{\bibinfo{volume}{133}}, \bibinfo{pages}{A171}
  (\bibinfo{year}{1964}).

\bibitem[{\citenamefont{Marzari and Vanderbilt}(1997)}]{Marzari97}
\bibinfo{author}{\bibfnamefont{N.}~\bibnamefont{Marzari}} \bibnamefont{and}
  \bibinfo{author}{\bibfnamefont{D.}~\bibnamefont{Vanderbilt}},
  \bibinfo{journal}{Phys. Rev. B} \textbf{\bibinfo{volume}{56}},
  \bibinfo{pages}{12847} (\bibinfo{year}{1997}).

\bibitem[{\citenamefont{Vanderbilt and King-Smith}(1993)}]{Vanderbilt93}
\bibinfo{author}{\bibfnamefont{D.}~\bibnamefont{Vanderbilt}} \bibnamefont{and}
  \bibinfo{author}{\bibfnamefont{R.~D.} \bibnamefont{King-Smith}},
  \bibinfo{journal}{Phys. Rev. B} \textbf{\bibinfo{volume}{48}},
  \bibinfo{pages}{4442} (\bibinfo{year}{1993}).

\bibitem[{\citenamefont{Lee et~al.}(2005)\citenamefont{Lee, Nardelli, and
  Marzari}}]{Marzari05}
\bibinfo{author}{\bibfnamefont{Y.-S.} \bibnamefont{Lee}},
  \bibinfo{author}{\bibfnamefont{M.~B.} \bibnamefont{Nardelli}},
  \bibnamefont{and} \bibinfo{author}{\bibfnamefont{N.}~\bibnamefont{Marzari}},
  \bibinfo{journal}{Phys. Rev. Lett.} \textbf{\bibinfo{volume}{95}},
  \bibinfo{pages}{076804} (\bibinfo{year}{2005}).

\bibitem[{\citenamefont{Kohn}(1959)}]{Kohn59}
\bibinfo{author}{\bibfnamefont{W.}~\bibnamefont{Kohn}}, \bibinfo{journal}{Phys.
  Rev.} \textbf{\bibinfo{volume}{115}}, \bibinfo{pages}{809}
  (\bibinfo{year}{1959}).

\bibitem[{\citenamefont{Stengel and Spaldin}(2006)}]{Stengel}
\bibinfo{author}{\bibfnamefont{M.}~\bibnamefont{Stengel}} \bibnamefont{and}
  \bibinfo{author}{\bibfnamefont{N.~A.}~\bibnamefont{Spaldin}},
  \bibinfo{journal}{Phys. Rev. B} \textbf{\bibinfo{volume}{73}},
  \bibinfo{pages}{075121} (\bibinfo{year}{2006}).

\bibitem[{\citenamefont{Rehr and Kohn}(1974)}]{RehrKohn}
\bibinfo{author}{\bibfnamefont{J.~J.}~\bibnamefont{Rehr}} \bibnamefont{and}
  \bibinfo{author}{\bibfnamefont{W.}~\bibnamefont{Kohn}},
  \bibinfo{journal}{Phys. Rev. B} \textbf{\bibinfo{volume}{10}},
  \bibinfo{pages}{448} (\bibinfo{year}{1974}).

\bibitem[{\citenamefont{Nenciu}(1983)}]{Nenciu83}
\bibinfo{author}{\bibfnamefont{G.}~\bibnamefont{Nenciu}},
  \bibinfo{journal}{Comm. Math. Phys.} \textbf{\bibinfo{volume}{91}},
  \bibinfo{pages}{81} (\bibinfo{year}{1983}).

\bibitem[{\citenamefont{Prodan and Kohn}(2005)}]{Prodan}
\bibinfo{author}{\bibfnamefont{E.}~\bibnamefont{Prodan}} \bibnamefont{and}
  \bibinfo{author}{\bibfnamefont{W.}~\bibnamefont{Kohn}},
  \bibinfo{journal}{Proc. Nat. Acad. Sci.} \textbf{\bibinfo{volume}{102}},
  \bibinfo{pages}{11635} (\bibinfo{year}{2005}).

\bibitem[{\citenamefont{Schnell et~al.}(2002)\citenamefont{Schnell, Czycholl,
  and Albers}}]{Schnell}
\bibinfo{author}{\bibfnamefont{I.}~\bibnamefont{Schnell}},
  \bibinfo{author}{\bibfnamefont{G.}~\bibnamefont{Czycholl}}, \bibnamefont{and}
  \bibinfo{author}{\bibfnamefont{R.~C.}~\bibnamefont{Albers}},
  \bibinfo{journal}{Phys. Rev. B} \textbf{\bibinfo{volume}{65}},
  \bibinfo{pages}{075103} (\bibinfo{year}{2002}).

\bibitem[{\citenamefont{Kim et~al.}(1995)\citenamefont{Kim, Mauri, and
  Galli}}]{KimMauri}
\bibinfo{author}{\bibfnamefont{J.}~\bibnamefont{Kim}},
  \bibinfo{author}{\bibfnamefont{F.}~\bibnamefont{Mauri}}, \bibnamefont{and}
  \bibinfo{author}{\bibfnamefont{G.}~\bibnamefont{Galli}},
  \bibinfo{journal}{Phys. Rev. B} \textbf{\bibinfo{volume}{52}},
  \bibinfo{pages}{1640} (\bibinfo{year}{1995}).

\bibitem[{\citenamefont{Nunes and Gonze}(2001)}]{Nunes}
\bibinfo{author}{\bibfnamefont{R.~W.}~\bibnamefont{Nunes}} \bibnamefont{and}
  \bibinfo{author}{\bibfnamefont{X.}~\bibnamefont{Gonze}},
  \bibinfo{journal}{Phys. Rev. B} \textbf{\bibinfo{volume}{63}},
  \bibinfo{pages}{155107} (\bibinfo{year}{2001}).

\bibitem[{\citenamefont{Sheng and Weng}(1995)}]{ShengWeng}
\bibinfo{author}{\bibfnamefont{D.~N.}~\bibnamefont{Sheng}} \bibnamefont{and}
  \bibinfo{author}{\bibfnamefont{Z.~Y.}~\bibnamefont{Weng}},
  \bibinfo{journal}{Phys. Rev. Lett.} \textbf{\bibinfo{volume}{75}},
  \bibinfo{pages}{2388} (\bibinfo{year}{1995}).

\bibitem[{\citenamefont{Walker and Wilkinson}(1995)}]{WalkerWilkinson}
\bibinfo{author}{\bibfnamefont{P.~N.}~\bibnamefont{Walker}} \bibnamefont{and}
  \bibinfo{author}{\bibfnamefont{M.}~\bibnamefont{Wilkinson}},
  \bibinfo{journal}{Phys. Rev. Lett.} \textbf{\bibinfo{volume}{74}},
  \bibinfo{pages}{4055} (\bibinfo{year}{1995}).

\bibitem[{\citenamefont{Canali et~al.}(2003)\citenamefont{Canali, Cehovin, and
  MacDonald}}]{Canali}
\bibinfo{author}{\bibfnamefont{C.~M.}~\bibnamefont{Canali}},
  \bibinfo{author}{\bibfnamefont{A.}~\bibnamefont{Cehovin}}, \bibnamefont{and}
  \bibinfo{author}{\bibfnamefont{A.~H.}~\bibnamefont{MacDonald}},
  \bibinfo{journal}{Phys. Rev. Lett.} \textbf{\bibinfo{volume}{91}},
  \bibinfo{pages}{046805} (\bibinfo{year}{2003}).

\bibitem[{\citenamefont{Hatsugai}(1993)}]{Hatsugai}
\bibinfo{author}{\bibfnamefont{Y.}~\bibnamefont{Hatsugai}},
  \bibinfo{journal}{Phys. Rev. Lett.} \textbf{\bibinfo{volume}{71}},
  \bibinfo{pages}{3697} (\bibinfo{year}{1993}).

\bibitem[{\citenamefont{Sheng et~al.}(2006)\citenamefont{Sheng, Weng, Sheng,
  and Haldane}}]{ShengHaldane}
\bibinfo{author}{\bibfnamefont{D.~N.}~\bibnamefont{Sheng}},
  \bibinfo{author}{\bibfnamefont{Z.~Y.}~\bibnamefont{Weng}},
  \bibinfo{author}{\bibfnamefont{L.}~\bibnamefont{Sheng}}, \bibnamefont{and}
  \bibinfo{author}{\bibfnamefont{F.~D.~M.}~\bibnamefont{Haldane}},
  \bibinfo{journal}{Phys. Rev. Lett.} \textbf{\bibinfo{volume}{97}},
  \bibinfo{pages}{036808} (\bibinfo{year}{2006}).

\bibitem[{\citenamefont{des Cloiseaux}(1964)}]{Cloizeaux1}
\bibinfo{author}{\bibfnamefont{J.}~\bibnamefont{des Cloiseaux}},
  \bibinfo{journal}{Phys. Rev.} \textbf{\bibinfo{volume}{135}},
  \bibinfo{pages}{A685} (\bibinfo{year}{1964}).

\bibitem[{\citenamefont{Strinati}(1978)}]{Strinati78}
\bibinfo{author}{\bibfnamefont{G.}~\bibnamefont{Strinati}},
  \bibinfo{journal}{Phys. Rev. B} \textbf{\bibinfo{volume}{18}},
  \bibinfo{pages}{4104} (\bibinfo{year}{1978}).

\bibitem[{\citenamefont{Katznelson}(1976)}]{Katznelson}
\bibinfo{author}{\bibfnamefont{Y.}~\bibnamefont{Katznelson}},
  \emph{\bibinfo{title}{An Introduction to Harmonic Analysis}},
  (\bibinfo{publisher}{Dover},
  \bibinfo{address}{New York}, \bibinfo{year}{1976}).

\bibitem[{\citenamefont{Bross}(1971)}]{Bross}
\bibinfo{author}{\bibfnamefont{H.}~\bibnamefont{Bross}},
  \bibinfo{journal}{Zeit. Phys.} \textbf{\bibinfo{volume}{243}},
  \bibinfo{pages}{311} (\bibinfo{year}{1971}).

\bibitem[{\citenamefont{Blount}(1962)}]{Blount}
\bibinfo{author}{\bibfnamefont{E.~I.}~\bibnamefont{Blount}}, in
  \emph{\bibinfo{booktitle}{Solid State Physics}}, edited by
  \bibinfo{editor}{\bibfnamefont{F.}~\bibnamefont{Seitz}} \bibnamefont{and}
  \bibinfo{editor}{\bibfnamefont{D.}~\bibnamefont{Turnbull}}
  (\bibinfo{publisher}{Academic Press}, \bibinfo{address}{New York},
  \bibinfo{year}{1962}), vol.~\bibinfo{volume}{13}, pp.
  \bibinfo{pages}{305--73}.

\bibitem[{\citenamefont{Kohn}(1973)}]{Kohn73}
\bibinfo{author}{\bibfnamefont{W.}~\bibnamefont{Kohn}}, \bibinfo{journal}{Phys.
  Rev. B} \textbf{\bibinfo{volume}{7}}, \bibinfo{pages}{4388}
  (\bibinfo{year}{1973}).

\bibitem[{\citenamefont{Teichler}(1971)}]{Teichler}
\bibinfo{author}{\bibfnamefont{H.}~\bibnamefont{Teichler}},
  \bibinfo{journal}{Phys. Stat. Sol. B} \textbf{\bibinfo{volume}{43}},
  \bibinfo{pages}{307} (\bibinfo{year}{1971}).

\bibitem[{\citenamefont{Chang}(1982)}]{ChangSi}
\bibinfo{author}{\bibfnamefont{Y.~C.}~\bibnamefont{Chang}},
  \bibinfo{journal}{Phys. Rev. B} \textbf{\bibinfo{volume}{25}},
  \bibinfo{pages}{605} (\bibinfo{year}{1982}).

\bibitem[{\citenamefont{Reed and Simon}(1978)}]{RSIV}
\bibinfo{author}{\bibfnamefont{M.}~\bibnamefont{Reed}} \bibnamefont{and}
  \bibinfo{author}{\bibfnamefont{B.}~\bibnamefont{Simon}},
  \emph{\bibinfo{title}{Methods of Modern Mathematical Physics}},
  vol.~\bibinfo{volume}{IV} (\bibinfo{publisher}{Academic Press},
  \bibinfo{address}{New York}, \bibinfo{year}{1978}).

\bibitem[{\citenamefont{Smogunov et~al.}(2004)\citenamefont{Smogunov, Corso,
  and Tosatti}}]{Smogunov}
\bibinfo{author}{\bibfnamefont{A.}~\bibnamefont{Smogunov}},
  \bibinfo{author}{\bibfnamefont{A.} \bibnamefont{Dal Corso}},
  \bibnamefont{and}
  \bibinfo{author}{\bibfnamefont{E.}~\bibnamefont{Tosatti}},
  \bibinfo{journal}{Phys. Rev. B} \textbf{\bibinfo{volume}{70}},
  \bibinfo{pages}{045417} (\bibinfo{year}{2004}).

\bibitem[{\citenamefont{Nenciu}(1991)}]{NenciuRMP}
\bibinfo{author}{\bibfnamefont{G.}~\bibnamefont{Nenciu}},
  \bibinfo{journal}{Rev. Mod. Phys.} \textbf{\bibinfo{volume}{63}},
  \bibinfo{pages}{91} (\bibinfo{year}{1991}).

\bibitem[{\citenamefont{Panati}(2006)}]{Panati}
\bibinfo{author}{\bibfnamefont{G.}~\bibnamefont{Panati}},
  \bibinfo{journal}{Ann. Henri Poincar{\'e}} \textbf{\bibinfo{volume}{8}},
  (\bibinfo{year}{2007}), 
  \bibinfo{note}{to be published, arXiv:math-ph/0601034}.

\bibitem[{\citenamefont{Goedecker}(1999)}]{Goedecker99}
\bibinfo{author}{\bibfnamefont{S.}~\bibnamefont{Goedecker}},
  \bibinfo{journal}{Rev. Mod. Phys.} \textbf{\bibinfo{volume}{71}},
  \bibinfo{pages}{1085} (\bibinfo{year}{1999}).

\bibitem[{\citenamefont{Strange}(1998)}]{Strange}
\bibinfo{author}{\bibfnamefont{P.}~\bibnamefont{Strange}},
  \emph{\bibinfo{title}{Relativistic Quantum Mechanics}}
  (\bibinfo{publisher}{Cambridge University Press},
  \bibinfo{address}{Cambridge}, \bibinfo{year}{1998}).

\bibitem[{\citenamefont{Krantz}(2001)}]{Krantz1}
\bibinfo{author}{\bibfnamefont{S.~G.}~\bibnamefont{Krantz}},
  \emph{\bibinfo{title}{Function Theory of Several Complex Variables}}
  (\bibinfo{publisher}{Am. Math. Soc.},
  \bibinfo{address}{Providence}, \bibinfo{year}{2001}).

\bibitem[{\citenamefont{Haldane}(2004)}]{Haldane}
\bibinfo{author}{\bibfnamefont{F.~D.~M.}~\bibnamefont{Haldane}},
  \bibinfo{journal}{Phys. Rev. Lett.} \textbf{\bibinfo{volume}{93}},
  \bibinfo{pages}{206602} (\bibinfo{year}{2004}).

\bibitem[{\citenamefont{Thouless}(1984)}]{Thouless}
\bibinfo{author}{\bibfnamefont{D.~J.}~\bibnamefont{Thouless}},
  \bibinfo{journal}{J. Phys. C} \textbf{\bibinfo{volume}{17}},
  \bibinfo{pages}{L325} (\bibinfo{year}{1984}).

\bibitem[{\citenamefont{Ohgushi et~al.}(2000)\citenamefont{Ohgushi, Murakami,
  and Nagaosa}}]{Ohgushi}
\bibinfo{author}{\bibfnamefont{K.}~\bibnamefont{Ohgushi}},
  \bibinfo{author}{\bibfnamefont{S.}~\bibnamefont{Murakami}}, \bibnamefont{and}
  \bibinfo{author}{\bibfnamefont{N.}~\bibnamefont{Nagaosa}},
  \bibinfo{journal}{Phys. Rev. B} \textbf{\bibinfo{volume}{62}},
  \bibinfo{pages}{6065} (\bibinfo{year}{2000}).

\bibitem[{\citenamefont{Resta}(1994)}]{RestaRMP}
\bibinfo{author}{\bibfnamefont{R.}~\bibnamefont{Resta}}, \bibinfo{journal}{Rev.
  Mod. Phys.} \textbf{\bibinfo{volume}{66}}, \bibinfo{pages}{899}
  (\bibinfo{year}{1994}).

\bibitem[{\citenamefont{Thonhauser et~al.}(2005)\citenamefont{Thonhauser,
  Ceresoli, Vanderbilt, and Resta}}]{Thonhauser}
\bibinfo{author}{\bibfnamefont{T.}~\bibnamefont{Thonhauser}},
  \bibinfo{author}{\bibfnamefont{D.}~\bibnamefont{Ceresoli}},
  \bibinfo{author}{\bibfnamefont{D.}~\bibnamefont{Vanderbilt}},
  \bibnamefont{and} \bibinfo{author}{\bibfnamefont{R.}~\bibnamefont{Resta}},
  \bibinfo{journal}{Phys. Rev. Lett.} \textbf{\bibinfo{volume}{95}},
  \bibinfo{pages}{137205} (\bibinfo{year}{2005}).

\end{thebibliography}
\end{document}